\documentstyle[12pt]{article}

\newcommand{\beq}{\begin{equation}}
\newcommand{\eeq}[1]{\label{#1} \end{equation}}
\newcommand{\beqn}{\begin{eqnarray}}
\newcommand{\eeqn}[1]{\label{#1} \end{eqnarray}}

\newcommand{\D}{{\cal{D}}}

\newcommand{\Z}{{\cal Z}}

\newcommand{\Tr}{{\rm Tr}}

\begin{document}

\vspace*{-2 cm}
\hfill\vbox{
\hbox{~~}
\hbox{\it La Plata-Th 99/05}
\hbox{hep-th/9905192} }

\vspace{1cm}

\centerline{\Large \bf Non-Abelian fractional quantum Hall states }

\centerline{\Large \bf and chiral coset conformal field theories}

\vspace{.8 cm}

\centerline{\bf
D.C.\ Cabra$^{a,b,}$\footnote{CONICET, Argentina}, \, E.\
Fradkin$^c$,
\, G.L.\ Rossini$^{a,1}$, }
\centerline{\bf and }
\centerline{\bf F.A.\ Schaposnik$^{a,}$\footnote{CICPBA, Argentina}}

\vspace{.8cm}

\centerline{\small\it $^a$Departamento de F\'{\i}sica,
Universidad Nacional de La Plata,}
\centerline{\small\it C.C. 67 (1900) La Plata, Argentina}
\centerline{\small\it $^b$
Facultad de Ingenier\'{\i}a, Universidad Nacional de Lomas de
Zamora,}
\centerline{\small\it Cno. de Cintura y Juan XXIII, (1832), Lomas de Zamora,
Argentina,}
\vspace{0.2cm}
\centerline{\small\it $^c$ Department of Physics, University of Illinois at
Urbana-Champaign}
\centerline{\small\it 1110 W.~Green St.~, Urbana, IL 61801, USA}

\centerline{ \date{}}

\vspace{2.2 cm}

\centerline{\bf Abstract} \vspace{0.1cm} {We propose an effective
Lagrangian for the low energy theory of the Pfaffian states
of the fractional quantum Hall effect in the bulk in terms of non-Abelian
Chern-Simons (CS) actions. Our approach exploits the connection
between the topological Chern-Simons theory and chiral conformal field
theories.
This construction can be used to describe a large class of non-Abelian FQH
states.}

\newpage

\pagenumbering{arabic}

\vspace{1cm}

\section{Introduction}
\label{sec:intro}


The Pffafian fractional quantum Hall (FQH) state has received much attention
recently, having been proposed as the leading candidate to describe the ground
state properties of the experimentally observed $\nu = 5/2$ FQH plateau
of a single layer two-dimensional electron gas (2DEG). It also bears a close
connection with the $\nu = 1/2$ state in bilayer systems. The Pfaffian states
\cite{read-moore,B} belong to a
family of FQH states whose excitations exhibit non-Abelian statistics.
All of these states involve some sort of pairing (or more generally
multi-particle bound states) among the electrons. \cite{rezayi-read}

The generalized $q-$Pfaffian wave function has the form
\begin{equation}
{\Psi_{\rm Pf}}\,\, =\,\,
{\rm Pf}\left(\frac{1}{{z_i} - {z_j}}\right)
\,\,\,{\prod_{i>j}}{\left({z_i} - {z_j}\right)^q}\,\,\,
{e^{-\frac{1}{4{\ell}_0^2} \sum |z_i|^2 }},
\label{GSwf}
\end{equation}
where $\ell_0$ is the cyclotron length, $\ell_0^2 = \hbar c/(eB)$.
This wave function, regarded as the ground state wavefunction of a system
of fully polarized fermions in a partially filled Landau level with a
local Hamiltonian, describes an incompressible ground state of the system.
Like in all quantum Hall states, all the excitations in the  bulk of the system
have a finite energy gap, and the only gapless
excitations reside at the physical boundary. Since these boundary or
edge excitations are gapless and, and as the external magnetic field breaks
time
reversal symmetry, the low energy Hilbert space of the system
is described by a chiral CFT. A salient feature of non-Abelian quantum
Hall states such as
the Pfaffian state is that their spectrum contains quasiparticles which exhibit
non-Abelian braiding statistics.

The incompressibility of the bulk state, the breaking of time
reversal invariance, and the existence of locally conserved
currents together imply that the low energy physics of the bulk
states should be describable in terms of a Chern-Simons gauge
theory with a suitably chosen gauge group \cite{wen-cs}. In this
way, the correspondence between the Hilbert space of the bulk FQH
state and its edge excitations reduces to the correspodence between
a Chern-Simons gauge theory and chiral CFT \cite{W1,witten2}. These
considerations led Wen to postulate that for any {\sl Abelian}
FQH state, the {\it edge chiral} CFT is determined by the
conformal block structure exhibited by the holomorphic factors of
the wave function (this being the origin of the concept of {\it
bulk} CFT) \cite{Wen,matrix}. Thus, for all the Abelian FQH states
there is a one-to-one correspondence between the operators that create
the excitations in the bulk and the spectrum of primary fields of the
associated chiral CFT of the edge states. Much of the current
understanding of the physics of the properties of the
edge states is based on this correspondence\cite{wen-cs}.

However, for the non-Abelian FQH states
the picture is still incomplete. Wave functions for non-Abelian FQH
states have been proposed\cite{read-moore,rezayi-read} and many of the
properties of the low energy
excitation spectra associated with these states have been studied
\cite{milovanovic,rezayi}. In a
number of cases the edge states of these non-Abelian FQH states are
fairly well understood\cite{milovanovic,nayak1}. For a number of
specific non-Abelian FQH states, notably the fermionic and the bosonic
Pfaffian states and their generalizations, effective Landau-like
theories for the bulk states have been proposed and their connections with
the pairing correlation of the wave functions and with the physics of the
CFT of their edge states have been worked out\cite{FNTW,FNS}. Recently,
Wen has also proposed a projective construction which is a promising
attempt for understanding many of the properties of these
states.\cite{Wen2}
However, a full classification
of the non-Abelian FQH states is lacking. In particular, the {\sl
natural} form of the effective theory in the bulk is not known, 
even though some of its main ingredients are known.

Drawing from the classification theory of the Abelian FQH states
in terms of generalized multicomponent Abelian Chern-Simons
theories \cite{matrix}, it is possible to infer what its natural
generalization to the  non-Abelian states should look like. In
this paper we propose an effective low energy theory for the bulk
Pfaffian FQH states. As expected this effective theory contains as
one of its main components, non-Abelian Chern-Simons theories with
a (quantized) coupling constant (known as the {\sl level}) $k >
1$. We will see, however, that the correct theory has several
sectors that need to be glued together carefully. Hence, the
effective theory  does {\sl not} reduce simply to a suitably
chosen non-Abelian Chern-Simons theory. By direct study of the
Pfaffian wave function, a number of authors have proposed specific
chiral edge theories for several non-Abelian states
\cite{milovanovic,rezayi,nayak1}. For the case at hand, the
Pfaffian state, the holomorphic Pfaffian factor in (\ref{GSwf})
corresponds to the conformal blocks of an Ising
CFT \cite{read-moore},  while the Laughlin factor corresponds to a
Rational Conformal Field Theory (RCFT) of a massless $U(1)$ chiral
boson with compactification radius $R = 1/\sqrt{q}$. The
non-Abelian structure of the Pfaffian state is encoded in the
Ising sector\cite{read-moore,nayak1}. In particular, the
corresponding edge chiral CFT is a direct product of an Ising
sector with symmetry ${\bf Z}_2$, and a $U(1)$ RCFT with compactification
radius $R=1/\sqrt{q}$,
\beq
Z_2
\otimes U(1)_{q} ~,
\eeq{cft}
with total central charge $c=1/2 + 1$.

It has been known for quite some time that
as a CFT, the Ising sector can also be described by an $SU(2)_2/U(1)$
coset CFT \cite{gko}. Thus, the CFT of the edge states can also be described by
a
CFT of a set of currents obeying the Kac-Moody symmetry (current algebra)
\beq
{\frac{SU(2)_2}{U(1)}} \otimes U(1)_q
\eeq{cft2}
The spectrum of a coset CFT of this type can be constructed
in terms of a (level $k$) $SU(2)_k$ gauged Wess-Zumino-Witten theory or,
equivalently, in terms of a chiral sector of a theory of constrained
fermionic currents (see {\it e.g.} \cite{CM} and references therein).
Coset constructions of the edge states of FQH states have also been
proposed \cite{FNTW,WZ}. These constructions are natural at the
level of the edge theory, and the cosets appear either as a result of the
need to project-out some sector of the theory, or by the need to change the
quantum numbers of the edge states ({\it i.\ e.\/}  the compactification
radius)
to get the right spectrum. In both instances the end result is
that some subgroup of the symmetry group is gauged. However, in some cases
gauging a subgroup at the edge leads to theories with broken
symmetries in the bulk, possibly due to a Higgs-like mechanism
\cite{FNTW,FNS} related with the existence of pairing correlations in the bulk
state.

In this paper we propose a simple effective theory for the bulk state
which has a natural connection with the known physics of the edge and at
the same time it represents a minimal departure from the conventional
Chern-Simons construction in the bulk. Guided by the structure of the
edge states we see that the problem of finding the
effective theory of the bulk is closely connected with the problem of
finding a Chern-Simons theory, at some level $k > 1$, whose
boundary is a coset CFT. Moreover, as the CFT is a chiral CFT with
constraints, the problem also involves finding the origin of the
constraints in the bulk physics and the connection with the physics
of pairing. Although recent work by several groups \cite{FNTW,FNS,Wen2,WZ}
has shown that, at least for a few specific cases, it is possible
to construct such theories, how general these constructions are is
not known.
In this work we will use the following alternative and simpler strategy.
Given that it is possible to determine the
structure of the CFT directly from the structure of the wave
function, and this CFT is known to be the chiral coset CFT $(SU(2)_2/U(1))
\otimes U(1)_q$, we will look for a natural bulk Chern-Simons
theory whose boundary CFT is this chiral coset. In addition, in
order to construct the spectrum of bulk excitations, we have to
supply a set of specific rules that will tell us how to put
together the representations of the different sectors of the theory
to make a physical excitation. In principle these rules should be a
consequence of the underlying physics of the bulk that causes the
system to have a Pfaffian state as its ground state.

The natural
candidate for such an effective theory should be a topological
quantum field theory on a manifold with boundary, supplied with a
set of appropriate boundary conditions. The boundary conditions
play two roles: (a) they {\it define} the actual {\it
dynamics} at the edge (and hence its energetics),
 and (b) they specify how the excitations are actually built up. This
approach is a natural generalization of the well known $U(1)$
Chern-Simons description of the Abelian FQH states (see {\sl e.\ g.\/}
\cite{Wen}). In this case it is well established that the full low
energy dynamics is described by a bulk (multicomponent) Abelian CS
theory, with edge modes described by a (set of) chiral $U(1)$ RCFT
of free bosons. Although this description works for generic Abelian FQH
states, it is not sufficient to reproduce the non-Abelian
structure of {\it e.\ g.\/ } the Ising sector of the Pfaffian state.
The main purpose of the present paper is to generalize
those previous schemes to the Pfaffian state and other non-Abelian
FQH systems.

 Here we use the general approach to non-Abelian CS
topological field theories on manifolds with boundaries, as
developed by Moore and Seiberg \cite{MS}, suitably extended to
include the boundary contitions relevant to the physics of the
Pfaffian state. Our approach should work for any non-Abelian FQH
state with a boundary CFT determined by a chiral current algebra
(as in the chiral coset CFT's).
We note, however, that there exist
non-Abelian FQH states which can be obtained by projecting a
generic multicomponent Abelian theory \cite{guille,cappelli} and
their CFT's are not generated by a current algebra. The relation
between these projected Abelian theories and the Pfaffian (and its
generalizations) is an interesting open question. This paper is
organized as follows. In section \ref{sec:connection} we review the
Chern-Simons-CFT within the path-integral framework of Moore and Seiberg
although with a new type of boundary conditions suitable to the physics of
quantum Hall systems. In particular we derive in detail the form of the
coset construction associated with these boundary conditions. In section
\ref{sec:bulk} we use the results of section \ref{sec:connection} to
determine the form of the effective theory of the bulk FQH Pfaffian
states. Here we show how the observables of the Pfaffian state are
represented in this theory and show that this theory is
consistent with the known properties of the Pfaffian states.
In Section \ref{sec:parafermions} we generalize of this
construction to the newly propsed parafermionic states.
Finally, in section \ref{sec:conclusions} we summarize our results 
and give a discussion of its validity.

\section{The Chern-Simons Conformal Field Theory connection in the path
integral framework}
\label{sec:connection}

We will summarize here the way to connect a CS theory on a
$3$-dimensional manifold $\Omega$ with boundary (say $D \times R$)
with a CFT on the boundary $\partial \Omega$ ($S_1 \times R$). The bulk
CS theory is a topological field theory (in the sense that the CS action
is independent of the metric of the manifold $\Omega$).
Therefore, on a manifold with a boundary, the definition of the theory must
include boundary conditions, and different conditions will in principle lead to
different
theories. We will adopt as a general criterion for boundary conditions that
surface terms should not appear in the equations of motion \cite{MS}.
However, unlike reference \cite{MS}, we will manage to control
the propagation velocity of the chiral modes at the boundary in
the spirit of \cite{Wen}, so as to end up with non-trivial boundary
dynamics.

The treatment of the bulk CS partition function that we present below has a
first step in which a component of the gauge field is
recasted as a Lagrange multiplier and integrated out. Then, the
delta function condition that arises is solved explicitly via an
adequate parameterization of the degrees of freedom, and all $3$-dimensional
integrals are
shown to depend only on the boundary fields, {\em i.\ e.\ }they are total
divergences or topological WZW terms.
We modify the procedure presented in \cite{MS} for
the case of chiral coset CFT's by choosing
suitable boundary conditions in the sense of getting the right
boundary dynamics, and establish the connection
between the effective theory
of the bulk and the corresponding chiral CFT at the boundary.
As a byproduct, we will find a set of rules that
characterizes the physical observables.

\subsection{Pure CS theories}

Consider the CS action on $\Omega$
\beq
S_{CS} =\frac{1}{4\pi}
\int_{\Omega} {d^3}x \epsilon^{\mu\nu\lambda} \Tr_{\hat G} (A_\mu
\partial_\nu A_\lambda +
\frac{2}{3} {A_\mu}{A_\nu}{A_\lambda} ),
\eeq{CS}
where the gauge field is taken in the Lie algebra of a group $G$. We use the
conventions
\beq
F_{\mu\nu}=\partial_{\mu}A_{\nu} -
\partial_{\nu}A_{\mu} + [ A_{\mu}, A_{\nu}],
\eeq{Fmunu}
\beq
A_\mu \;=\; A_\mu^a \tau_a\,,\, A_\mu^a \in \Re
\eeq{conventions}
with $\tau_a^\dagger = -\tau_a$, $[\tau_a , \tau_b] \,=\,f_{abc} \,\tau_c$ and
$\Tr_{\hat G}(\tau_a \tau_b)= -\frac{1}{2}\delta_{ab}$ for the non-Abelian
$SU(N)$ case.
In components the CS action reads
\beq
S_{CS} =-\frac{1}{8\pi}
\int_{\Omega} {d^3}x \epsilon^{\mu\nu\lambda}  (A^a_\mu
\partial_\nu A^a_\lambda +
\frac{1}{3} f_{abc}{A^a_\mu}{A^b_\nu}{A^c_\lambda} ),
\eeq{CScomp}

The variation of the action is
\beq
\delta S_{CS} = -\frac{1}{8\pi} \int_{\Omega}
{d^3}x\,\epsilon^{\mu\nu\lambda} \delta {A_\mu^a} F_{\nu\lambda}^a
+ \frac{1}{8\pi} \int_{\partial\Omega} dS_{\mu}\,
\epsilon^{\mu\nu\lambda} {A_{\nu}^a}\delta{A_{\lambda}^a},
\eeq{deltaCS}
where $ dS_{\mu}$ is a surface element normal to
$\partial\Omega$. The boundary conditions should be such that the
surface term vanishes for any allowed variation of the gauge field. In
our geometry $D \times R$ different
possibilities are to set $A_t=0$, or $A_x=0$, or any linear combination
such as $A_t+v A_x=0$ (being $\partial\Omega= S_1 \times R$, we use $x$
for the compact direction $S_1$ and $t$ for the real line $R$).

Let us first consider $A_t=0$ on $\partial\Omega$ \cite{Elitzur}.
In this case one can write
\beq
S_{CS}= - \frac{1}{4\pi} \int_{\Omega}
{d^3}x \epsilon^{ij}A_{i}^a
\partial_t A_{j}^a +
\frac{1}{2\pi} \int_{\Omega} {d^3}x \epsilon^ {ij}A_t^a
(\partial_{i} A_{j}^a+ f_{abc} A_{i}^b A_{j}^c),
\eeq{CS2}
where now $i,j$ indicate coordinates $x,y$ on $D$.
Surface terms have been discarded {\em using the boundary
condition}.

The partition function
\beq
\Z = \int  \D A_{\mu} \exp (iS_{CS})
\eeq{ZCS}
can now be easily transformed in the following way: $A_t$ is
integrated out, imposing the condition $F_{ij}=0$. Then the only
contributing configurations are explicitely parameterized as a 2D
pure gauge,
\beq
A_{i}= -\partial_{i} g g^{-1}, ~~~~~~ (i=x,y)
\eeq{g}
It is worth noting that the Jacobians arising from the change of
variables $A_1 \to g$ cancel out exactly with the one coming from
the delta functional written in terms of the new fields.
The remaining effective action reads
\beq
S_{eff} = -\frac{1}{4\pi}\int_{\Omega} d^{3}x\epsilon^{ij} \Tr_{\hat G}
\left( \partial_i gg^{-1}\partial_0(\partial_j gg^{-1}) \right).
\eeq{Srest}
After some algebra, the partition function reads
\beq
{\cal Z}= \int {\cal D} g \exp(i\tilde{S}_{WZW}^{Ch}[g])
\eeq{ZCSa}
where $\tilde{S}_{WZW}^{Ch}$ is a kind of chiral Wess-Zumino-Witten (WZW)
action,
\begin{eqnarray}
\tilde{S}_{WZW}^{Ch}[g]&=&\frac{1}{4\pi} \int_{\partial\Omega}
d^2x~\Tr_{\hat G}\left(g^{-1}\partial_x gg^{-1}
\partial_t g\right) -
\nonumber \\
& & \frac{1}{12\pi} \int_{\Omega} d^{3}x~\epsilon^{\mu\nu\alpha}
\Tr_{\hat G}\left(g^{-1}\partial_{\mu} g g^{-1}\partial_{\nu} g
g^{-1}\partial_{\alpha} g\right).
\label{cwzw1}
\end{eqnarray}
However, this result presented in \cite{Elitzur} is not
satisfactory for our present interest in the sense that the action
(\ref{cwzw1}) has no propagating degrees of freedom. In fact, the
quadratic term gives just the simplectic structure of the 2D
theory, no kinetic term is present in the action. This structure is a
consequence of the fact the the Chern-Simons theory is topological (in
the sense of being invariant under transformations of the metric). The
gauge choice $A_t=0$ is consistent with general covariance and thus the
effective theory has no dynamics. However, in a general gauge and/or
for a general choice of boundary conditions general covariance will be
broken. Clearly, different boundary conditions lead to different
dynamics of the effective theory. Which choice is physically correct
cannot be determined from Chern-Simons theory alone.


\subsubsection{Modification of the boundary conditions}


Following the approach introduced by Wen in the abelian theories\cite{Wen}, we
will
now change the theory by modifying the boundary conditions to
$A_t+v A_x=0$. The previous computation can be copy cut if we define new
coordinates
on $\Omega$ as
\beq
\begin{array}{l}
\tilde{x}=x+vt\\
\tilde{y}=y\\
\tilde{t}=t
\end{array}
\eeq{newx}
so that the covariant vectors change as
\beq
\begin{array}{l}
\tilde{A}_x=A_x\\
\tilde{A}_y=A_y\\
\tilde{A}_t=A_t-vA_x
\end{array}
\eeq{newA}
The action (\ref{CS}) is invariant under this change of
coordinates, so that all the previous computations proceed in the
same way, writing $\tilde{A}$ instead of $A$. At the end of the
procedure we get as effective action the chiral WZW action as
studied in \cite{Stone}:
\begin{eqnarray}
S_{WZW}^{Ch}&=&\frac{1}{4\pi} \int_{\partial\Omega}
d^2x~\Tr_{\hat G}\left(g^{-1}\partial_x gg^{-1}
(\partial_t -v \partial_x) g\right) -
\nonumber \\
& & \frac{1}{12\pi} \int_{\Omega} d^{3}x~\epsilon^{\mu\nu\alpha}
\Tr_{\hat G}\left(g^{-1}\partial_{\mu} g g^{-1}\partial_{\nu} g
g^{-1}\partial_{\alpha} g\right),
\label{cwzw}
\end{eqnarray}
which describes propagating edge excitations as desired.

For later convenience, let us write out the particular Abelian
($\hat{G}=\hat{U(1)}$) form of eq.(\ref{cwzw}),
\beq
S_{\hat{U(1)}}^{Ch}=-\frac{1}{4\pi} \int_{\partial\Omega}
d^2x~\partial_x \phi
(\partial_t -v \partial_x) \phi,
\eeq{chboson}
which is of course nothing but the usual chiral boson action
\cite{FloJackiw,Siegel}.


\subsection{The CS-CFT connection for Coset theories}

As we pointed out in the introduction, the Ising CFT of interest in this
paper can be formulated as a
$SU(2)_2/U(1)$ coset CFT.
For convenience, we will extend the previous treatment to the case in which the
boundary theory is a coset ${\hat G}_{k_G}/{\hat H}_{k_H}$, $k_G=lk_H$, with
$l$ being the index of the embedding of $\hat H$ in $\hat G$. We will follow
the approach in
\cite{MS} as well as a suitable modification in the boundary conditions
so as to study the physical realization of the Pfaffian state
\cite{FNTW}.

We begin with the action of two gauge fields $A$ and $B$ in the
bulk in $2+1$ dimenions, where $A$ is
in the Lie algebra of ${\hat G}_{k_G}$ and $B$ is in the Lie algebra of
${\hat H}_{k_H}$. The action $S$ is just the difference
of the CS action of ${\hat G}_{k_G}$ and that of ${\hat H}_{k_H}$,
\beq
S=k_GS_{CS}[A,\hat G]-k_HS_{CS}[B,\hat H],
\eeq{coset}
supplied with the boundary conditions
\begin{eqnarray}
{\cal P}_{\hat H^{\perp}} A_t =&&\!\!\!\! 0, \\
k_G {\cal P}_{\hat H} A_i -k_H
B_i = &&\!\!\!\!0, \ \ (i=x,t) \ \ {\rm on  }\ \partial \Omega
\label{bc}
\end{eqnarray}
These boundary conditions do not break general covariance and
consequently do not give dynamis to the effective theory.
We consider again $\Omega = D\times R$ and hence $\partial
\Omega =
S_1\times R$.
${\cal P}_{\hat H}$ and ${\cal P}_{\hat H^{\perp}}$ denote
the projectors
on the Lie algebra $\hat H$ and its orthogonal complement
respectively. These boundary conditions satisfy the
general
criterion of having no boundary terms in the equations of
motion,
and allow to write each CS term in the form of
eq.(\ref{CS2}) as
boundary terms arising from the integration by parts vanish.

After integrating out $A_t$ and $B_t$ the partition function takes the form
\beqn
{\cal Z}&=&\int DA_i DB_i \delta \left[ \epsilon^{ij}
F_{ij}[A]\right]
\delta \left[ \epsilon^{ij} F_{ij}[B]\right]
\nonumber \\
& & \exp  -\frac{i}{4\pi}
\int_\Omega d^3x \epsilon^{ij}\left(k_G \Tr_{\hat G} (A_i \partial_t A_j) -
k_H\Tr_{\hat H} (B_i \partial_t B_j)
\right),
\eeqn{pf}
where $i,j$ are the indices corresponding to the coordinates on $D$.

The zero curvature constraints are easily solved on the simply connected
2D manifold $D$ in terms of group valued fields $g \in G$, $h \in H$
\beq
A_{i}= -\partial_{i} g g^{-1},\ \ \ \
B_{i}= -\partial_{i} h h^{-1}. ~~~~~~ (i=x,y)
\eeq{g2}
The boundary conditions on the $x$ components have to be explicitly
implemented. This is done by means of a Lagrange multiplier $\lambda \in
\hat H$
\beq
\int D\lambda \exp \left( i\int d^2 x
\Tr_{\hat H}\left(\lambda(k_G\partial_x g g^{-1} -
k_H\partial_x h h^{-1}) \right)\right).
\eeq{lm}

Putting all these things together we get
\beqn
{\cal Z}=\int Dg Dh D\lambda \exp \left(
ik_GS_{WZW}^{Ch}[g] - ik_H S_{WZW}^{Ch}[h] +
\right.
\nonumber \\
\left. \frac{ik}{2\pi}
\int d^2 x \Tr_{\hat H} \left(\lambda(k_G\partial_x g g^{-1} -
k_H\partial_x h h^{-1}) \right)
\right) .
\eeqn{pf2}
Now, being $k_G=l k_H$, we change variables to
\beq
U=gh^{-1}, \ \ \ \lambda = C_t, \ \ \ - \partial_x h
h^{-1} = \frac{1}{\sqrt l} C_x ,
\eeq{chv}
where $C_x, C_t$ are gauge fields in $\hat{H}$.
Using the Polyakov-Wiegmann identity we arrive to an effective action in
the form of a Chiral gauged WZW action
\begin{eqnarray}
\lefteqn{k_G\tilde{S}_{WZW}^{Ch, gauged}[U,C_i]= k_G\tilde{S}_{WZW}^{Ch}[U]
}&& \nonumber \\
&&+ \frac{k_G}{2\pi} \int_{\partial\Omega} dt\ dx\  Tr_{\hat H}
(-C_xU^{-1}\partial_t U + C_t \partial_x U U^{-1} - C_t U C_x
U^{-1}+ C_t C_x)
\nonumber \\
&&
\label{GWZWMS}
\end{eqnarray}
for the fields $U\in G$, $C_i \in \hat H$
(the functional integral over $h$ is factored out as the volume of
the group $H$). This effective action corresponds to the coset theory
${\hat G}_{k_G}/{\hat H}_{k_G}$.

Here again, we use the tilde notation $\tilde{S}_{WZW}^{Ch,
gauged}$ in order to emphasize that this is a kind of chiral
theory with no propagating degrees of freedom. Once again, we will
modify minimally the boundary conditions in order to break general
covariance, thus giving dynamics to the effective theory.


\subsubsection{Modification of the boundary conditions
for the coset theory}


We can easily modify the spectrum of the theory and introduce
chiral propagating excitations by a suitable modification of the
boundary conditions for the CS fields. In fact, it is enough to use
the coordinates in eqs.(\ref{newx}, \ref{newA}) and impose the
boundary conditions (\ref{bc}) for the tilde components of the
gauge fields. The modified boundary conditions read
\begin{eqnarray}
{\cal P}_{\hat H^{\perp}} (A_t+vA_x) =&&\!\!\! 0, \\
 k_G {\cal P}_{\hat H} A_i -k_H B_i =&&\!\!\!  0, \ \ (i=x,t)
\ \ {\rm on  }\ \partial \Omega
\label{bc'}
\end{eqnarray}
The resulting effective action reads
\begin{eqnarray}
\lefteqn{k_G{S}_{WZW}^{Ch, gauged}[U,C_i]= k_G{S}_{WZW}^{Ch}[U]}&&
\nonumber \\
&&+ \frac{k_G}{2\pi} \int_{\partial\Omega} dt\ dx\ Tr_{\hat H}
(-C_{\tilde{x}}U^{-1}\partial_{\tilde{t}} U + C_{\tilde{t}}
\partial_{\tilde{x}}U U^{-1} - C_{\tilde{t}} U
C_{\tilde{x}} U^{-1}+ C_{\tilde{t}} C_{\tilde{x}}) \nonumber \\
&&
\label{GWZWWen}
\end{eqnarray}
The kinetic term here insures that the action in Eq.\ (\ref{GWZWWen})
describes a propagating chiral CFT for the ${\hat
G}_{k_G}/{\hat H}_{k_G}$ coset theory.

We conclude this section by describing the content of physical
observables in the coset CS theory: the gauge group of the coset CS
theory is not simply $G \times H$ due to the imposed boundary
conditions.
In fact, the common center of $G$ and $H$ has to be mod
out. Then,
the observables in the coset CS theory are built up as products of
Wilson loop operators in representations $\Lambda$ in $G$ and
$\lambda$ in $H$ satisfying the following rules \cite{MS}:
\begin{enumerate}
\item
both representations should transform in the same way under the
common center so that the product of Wilson loops is invariant
under its action.
\item
representations related by the spectral flow associated to the
center should be identified.
\end{enumerate}
Hence, gluing the theories at the boundary, as is done through the
boundary conditions (\ref{bc}), insures that only the physical
representations are allowed to appear in the bulk, and that the
allowed representations of the bulk states are in one-to-one
correspondence with the allowed states at the edge. This property
insures that the observables of the coset CS theory correspond to
the correct integrable representations of the coset CFT \cite{MS}.

\section{Bulk theory for the $\nu = 1/q$ Pffafian FQHE state}
\label{sec:bulk}

The mathematical setting described above can be now readily applied
to construct an effective action for the $\nu = 1/q$ Pffafian FQHE
state ($q$ even). The {\sl action} of the effective theory
 is given by the sum of two terms,
\beq
S_{Pfaff}^{bulk}= S_{Z_2}+S_{U(1)_q},
\eeq{Sproposed}
corresponding to the Ising and $U(1)$ sectors respectively. However, the
Hilbert space of the theory is smaller since the only allowed states are
those that at the boundary satisfy the gluing conditions. In particular
only the representations that are invariant under the simultaneous
actions of the center $Z_2$ of both groups survive (see below).

The action for the Ising sector ($SU(2)_2/U(1)$)  is given by eq.
(\ref{GWZWWen}), with $k_G=k_H=2$, $G=SU(2)$ and $H=U(1)$
\beq
S_{Z_2}= 2S_{CS}[A,\hat{SU(2)}]-2S_{CS}[B,\hat{U(1)}],
\eeq{coset2}
together with the boundary conditions (\ref{bc'}).

The effective action for the $U(1)$ sector simply reads
\beq
S_{U(1)_q} = q S_{CS}[C,\hat{U(1)}],
\eeq{uuno}
with the boundary conditions $C_t+v C_x=0$, which insure the chiral
boson properties for the charge degree of freedom on the edge and do not couple
the $C$-field to the Ising sector.

The main point in this paper is that the bulk partition function
\beq
\Z_{bulk} =
\int \D A_{\mu} \D B_{\mu}\D C_{\mu} \exp(-S_{Pfaff}^{bulk})
\eeq{Zbulk}
with the imposed boundary conditions can be identified, as shown
throughout this work, with that of the edge theory
\beq
\Z_{edge}
= \int \D U \D C_{i} \D \phi  \exp(-2S_{WZW}^{Ch, gauged}[U,C_i]
-q S_{\hat{U}(1)}^{Ch}[\phi]).
\eeq{Zbulk2}
Moreover, we are in conditions to give a detailed description of
the observables of the proposed theory:
\begin{enumerate}
\item
For the case of the $SU(2)_2/U(1)$ coset chiral edge CFT, the
gauge group of the coset CS theory is
\beq
\frac{SU(2) \times U(1)}{Z_2}.
\eeq{CSgauge}
The CS observables correspond to Wilson loop operators in the
integrable representations of the loop group \cite{W1}.
In the case of $SU(2)_2$ they are labeled by the spin $j$ and restricted
to $\Lambda_0, \Lambda_{1/2}, \Lambda_1,$ \cite{Segal}.
For the $U(1)$ CS, the corresponding representations $\lambda_r$ are
labeled by the $U(1)$ charge which is restricted to the set
$r=0,1,2$ \cite{MS}.
Being $Z_2$ the common center for these
groups, transformations by the center just mean parity and the
rule (1) matches together $(\Lambda_0 , \lambda_0)$,
$(\Lambda_1 , \lambda_0)$ and $(\Lambda_{1/2} , \lambda_1)$, the
other pairs being redundant by rule (2) ({\it i.e.} the
associated spectral flow relates $(\Lambda_{j}
, \lambda_r) \rightarrow (\Lambda_{1-j} , \lambda_{r+2})$ ).
This leads to the correct representations for the edge chiral CFT,
which respectively correspond to the identity operator, the
Majorana fermion ($\psi_M$) and the spin operator ($\sigma$).

\item
In the case of a direct-product CS theory, (such as the present
$Z_2 \times U(1)_q$ in the Pfaffian), the boundary conditions do
not relate observables, and we still have to combine together the
integrable representations of the $Z_2$ sector with those of the
level $2q$ $U(1)$ sector to build up the integrable representations
of the whole theory. All different possible products of the
integrable representations of the sectors should correspond to the
allowed ones in the direct-product CS theory.
However, as discussed in \cite{WenPf}, not all the operators
in the direct product theory are physical.
In particular, for $q=2$, the electron operator is given by
$\psi_M \times \exp (i \sqrt{q} \phi)$ and the quasiparticle
operator by $\sigma \times \exp (i \phi / (2\sqrt{q}) )$
\cite{read-moore,WenPf}.

\end{enumerate}

The topological properties of the proposed effective action
characterize completely and faithfully the universality class corresponding to
the
Pfaffian state. In fact, the theory predicts the same topological
properties as that proposed in \cite{FNTW}. In particular it gives the correct
 the ground state degeneracy on the torus, the degeneracy of the quasihole
states and the correct non-Abelian braiding statistics. This last result
is a direct consequence of
the fact that the non-trivial non-Abelian structure is encoded in
the Ising factor, which inherits its structure from the $SU(2)_2$
theory from which it is derived.

\section{Parafermionic states}
\label{sec:parafermions}

We can also apply our method to construct the effective action
for the parafermionic states, recently proposed in \cite{Pf},
which are described by the wave function
\beq
\Psi^M_{Para} (z_1,z_2,..,z_N) =
\langle \psi_1 (z_1)\psi_1(z_2)...\psi_1 (z_N)
\rangle \prod_{i<j}(z_i-z_j)^{M+2/k},
\eeq{gsparaf}
where $\psi_1$ stands for the basic parafermion field in
the $Z_k$ CFT, $M$ odd (even) corresponds to a fermionic (bosonic) state
and the filling factor is given by $\nu = 1/(M+2/k)$.

~From this wave function, (assuming as usual that bulk and boundary
CFT's coincide), one can read
off the boundary chiral CFT, which is given by the direct product

\beq
Z_k \otimes U(1)_{N/2} ,
\eeq{paraf}
where $Z_k$ stands for the parafermionic CFT \cite{FK,ZF} and
the bosonic RCFT has level $N=k(kM+2)/2$.

As for the Ising boundary CFT, we use the coset construction $Z_k
= SU(2)_k/U(1)$ for the parafermionic sector, and then the whole
previous analysis applies, the only change being in the level $k$ of
the $SU(2)$ WZW sector. As discussed in \cite{read-moore}, rules (1) and
(2) described above reproduce the right operator content of the
$Z_k$ parafermionic CFT. In particular, the basic parafermion
operator $\psi_1$, (which is the generalization of the Majorana fermion
for $k > 2$), and the basic spin field $\sigma_1$ are built up
as $(\Lambda_1 , \lambda_0)$ and $(\Lambda_{1/2} , \lambda_1)$
respectively.

Then, the electron and quasiparticle operators proposed in \cite{Pf} are
identified in our notation with $\psi_1 \times \exp (i \sqrt{M+2/k}
\phi)$ and the quasiparticle operator by $\sigma_1 \times \exp (i
\phi / (k\sqrt{M+2/k}) )$.

\section{Conclusions}
\label{sec:conclusions}

In summary, in this paper we proposed a general method to
construct effective actions for non-Abelian FQHE states, following
mainly the techniques developed in \cite{MS} for CS topological
field theories on manifold with boundaries. We showed that
suitably modified boundary conditions lead to the correct edge
description for the chiral (gapless) physical propagating modes.
We also study the issue of the allowed representations in such
theories which provide a one-to-one correspondence between bulk
and edge observables. In particular, we have seen that for a given
action describing the physics of the edge, {\em e.\ g.\ }in the case of
the $q$-Pfaffian state which action is given by
\beq
S_{Pfaffian}^{edge} = S_{Z_2} + S_{U(1)_q},
\eeq{fin.1}
supplied with a suitable set of boundary conditions that insure that the
gauge group of the coset theory is actually $(SU(2) \times U(1))/Z_2$,
we can construct the corresponding bulk effective field theory action,
which is given by
\beq
S_{Pfaffian}^{bulk} = 2S_{CS}[A,\hat{SU(2)}]-
2S_{CS}[B,\hat{U(1)}] + q S_{CS}[C,\hat{U(1)}].
\eeq{fin.2}
We have proven this connection through the equality of the
corresponding partition functions
\beq
\Z_{bulk} \equiv \Z_{edge}
\eeq{fin.3}
and we have furthermore shown a one-to-one
correspondence between observables (physical operators) in the
edge and in the bulk. The proposed bulk theory satisfies all the
requirements needed to describe the physics of the $q$-Pfaffian state, in
the following sense:
\begin{enumerate}
\item
The boundary chiral CFT corresponds to the direct product
$\frac{SU(2)_2}{U(1)} \otimes U(1)_q$
\item
It has the right number of propagating degrees of freedom as
indicated by the central charge $c=3/2$.
\item
It has the right content of physical observables.
\end{enumerate}
One of the main advantages of the present approach is that it can
be extended to other non-Abelian FQHE states, such as those
recently proposed in references
\cite{Wen2} and \cite{Sch}. The
connection between the proposed effective action with those given
by previous authors \cite{FNTW}, \cite{Wen2},  and the microscopic
origin of such an effective action are open points that remain to
be studied.

\vspace{5mm}

{\em Acknowledgments:} We would like to thank A.\ Lugo for helpful
discussions. E.F.\ thanks C.\ Nayak for many insightful remarks.
E.F.\
was a visitor at the Departamento de F{\'\i}sica of the Universidad
de La Plata when this work was started. D.C.C.\ thanks the
Physikalisches Institut der Universit\"at Bonn for their kind
hospitality where part of this work has been done. This work is
partially supported by CICBA, CONICET (PIP 4330/96), ANPCyT (grants
No.\ 03-00000-02249 and No.\ 03-00000-02285), Fundaci\'on
Antorchas, Argentina, and by the US National Science Foundation,
grant No.\ NSF-DMR98-17941 at the University of Illinois.

\end{document}